\documentclass{webofc}
\usepackage[varg]{txfonts}
\usepackage[utf8]{inputenc}
\usepackage{graphicx,amsmath,amssymb,bm}
\usepackage{hyperref,url}
\hypersetup{colorlinks=true,citecolor=blue,urlcolor=blue,linkcolor=blue}
\newcommand{\as}{\alpha_s}
\newcommand{\LQCD}{\Lambda_{\mathrm{QCD}}}
\newcommand{\dd}{\mathrm{d}}
\begin{document}
\title{The QCD Trace Anomaly in Pion Gravitational Form Factors with Sudakov Resummation}
\author{\firstname{Claudio} \lastname{Corianò}\inst{1,2} \and
\firstname{Dario} \lastname{Melle}\inst{1}\fnsep\thanks{Presenter} \and
\firstname{Leonardo} \lastname{Torcellini}\inst{1}}
\institute{Dipartimento di Matematica e Fisica, Università del Salento and INFN Sezione di Lecce,
Via Arnesano, 73100 Lecce, Italy; National Centre for HPC, Big Data and Quantum Computing
\and CNR Nanotec, Via Monteroni, 73100 Lecce, Italy}
\abstract{We discuss the pion gravitational form factors at intermediate and large spacelike momentum transfer in a perturbative-QCD factorization framework. The short-distance kernel is supplemented by the one-loop non-Abelian $TJJ$ correlator, where the energy--momentum tensor couples to two off-shell gluons. Its trace sector contains the QCD conformal anomaly and admits a scalar, dilaton-like pole interpretation. A transverse-momentum-dependent pion wave function, combined with Sudakov resummation and a Gaussian intrinsic-transverse-momentum profile, suppresses endpoints and large transverse separations. The projection on the pion form factors exhibits a characteristic hierarchy: the isolated anomaly cancels in the momentum form factor $A_\pi$, remains important in the mechanical form factor $D_\pi$, and dominates the $TJJ$ correction to the trace form factor. The result identifies a beta-function-controlled short-distance contribution to pion mechanical structure and clarifies the relation between the perturbative anomaly pole and a correlated scalar partonic channel.}
\maketitle

\section{Introduction}
\label{sec:introduction}
The matrix element of the QCD energy--momentum tensor (EMT) provides a unified description of how energy, momentum and mechanical forces are distributed inside a hadron. Its invariant amplitudes, conventionally called gravitational form factors (GFFs), are not measured by coupling a hadron to a macroscopic gravitational field. Instead, they enter moments of generalized parton distributions and may be constrained in hard exclusive reactions. This connection has led to the first phenomenological determinations of pressure and shear distributions in the proton and to growing interest in the corresponding pion observables \cite{Burkert:2018bqq,Kumano:2017lhr}. Lattice QCD has meanwhile provided direct determinations of pion quark and gluon GFFs over a useful range of spacelike momentum transfer \cite{Hackett:2023pyn}.

At large momentum transfer, hard exclusive amplitudes admit a factorized treatment in which a perturbatively calculable partonic kernel is convoluted with universal hadronic distribution amplitudes. The same logic that underlies the perturbative pion electromagnetic form factor can be applied to the matrix element of the EMT \cite{Lepage:1980fj,Tong:2022zax}. The gravitational case, however, has an additional structural ingredient. The renormalized trace of the QCD EMT is nonzero even in the chiral limit,
\begin{equation}
T^\mu_{\ \mu}=\frac{\beta(g_s)}{2g_s}F^a_{\rho\sigma}F^{a\rho\sigma}
+\sum_f(1+\gamma_m)m_f\bar\psi_f\psi_f,
\label{eq:trace-anomaly}
\end{equation}
up to BRST-exact and equation-of-motion operators. The scale anomaly can therefore enter an exclusive hard kernel through a stress-tensor insertion on gluonic lines.

The relevant object is the non-Abelian three-point function $\langle TJJ\rangle$, with one EMT and two color-current or gluonic insertions. Renormalization of this vertex produces a scalar trace form factor fixed by the QCD beta function. In momentum space it displays the anomaly-pole structure that is commonly represented as a perturbative dilaton exchange \cite{Armillis:2010qk,Coriano:2024qbr}. The word “dilaton” is used here in this precise sense: it is not a new elementary scalar added to QCD, but the coherent scalar projection of a renormalized composite correlator.

This contribution summarizes the analysis of Ref.~\cite{Coriano:2026gff}, emphasizing the elements most relevant for the QCD@Work setting. We first define the pion GFFs and their hard-scattering factorization, then describe how the trace sector of the off-shell $TJJ$ vertex is matched onto the pion amplitude. We subsequently introduce transverse momentum and Sudakov resummation, which are essential for controlling endpoint and large-distance configurations at finite $Q^2$. Finally, we discuss the distinct projections onto $A_\pi$, $D_\pi$ and the trace form factor, and the dispersive interpretation of the anomaly as a correlated scalar channel.

\section{Pion form factors and hard factorization}
\label{sec:factorization}
For a spin-zero hadron, conservation and Lorentz covariance reduce the matrix element of the total EMT to two form factors,
\begin{equation}
\langle \pi(P_2)|T^{\mu\nu}|\pi(P_1)\rangle
=2P^\mu P^\nu A_\pi(q^2)
+\frac{1}{2}\left(q^\mu q^\nu-g^{\mu\nu}q^2\right)D_\pi(q^2),
\label{eq:gff-decomposition}
\end{equation}
where $P=(P_1+P_2)/2$, $q=P_2-P_1$ and $Q^2=-q^2>0$. The normalization $A_\pi(0)=1$ expresses the momentum sum rule for the full EMT. The $D$-term contains information on the internal pressure and shear forces. Taking the trace gives
\begin{align}
\Theta_\pi(Q^2)&\equiv g_{\mu\nu}\langle \pi(P_2)|T^{\mu\nu}|\pi(P_1)\rangle \nonumber\\
&=2\left(m_\pi^2-\frac{q^2}{4}\right)A_\pi(q^2)-\frac{3}{2}q^2D_\pi(q^2).
\label{eq:trace-form-factor}
\end{align}
In the hard chiral limit this becomes
\begin{equation}
\Theta_\pi(Q^2)\simeq \frac{Q^2}{2}\left[A_\pi(-Q^2)+3D_\pi(-Q^2)\right].
\label{eq:hard-trace}
\end{equation}
Equation~\eqref{eq:hard-trace} already shows why a trace insertion need not shift the two GFFs equally.

At $Q^2\gg \LQCD^2,m_\pi^2$, the leading Fock component of the pion consists of a compact quark--antiquark pair. Introducing the longitudinal fractions $x$ and $y$ for the incoming and outgoing pion, the leading-power matrix element can be written schematically as
\begin{align}
\langle \pi(P_2)|T^{\mu\nu}|\pi(P_1)\rangle
=\frac{i f_\pi^2 C_F g_s^2}{N_c}
\int_0^1\!\dd x\int_0^1\!\dd y\,
\phi_\pi(x,\mu)K^{\mu\nu}(x,y,Q,\mu)\phi_\pi(y,\mu).
\label{eq:collinear-factorization}
\end{align}
The hard momentum transferred to one constituent is redistributed through a gluon with virtuality of order $Q^2$. At the lowest nonvanishing order the EMT couples directly to a quark or to the exchanged gluon, while the long-distance binding information is encoded in the leading-twist distribution amplitude $\phi_\pi$.

For matching purposes the hard kernel may be expanded in the five-tensor basis
\begin{equation}
K^{\mu\nu}=F_1P_1^\mu P_2^\nu+F_2P_2^\mu P_1^\nu
+F_3P_1^\mu P_1^\nu+F_4P_2^\mu P_2^\nu+F_5g^{\mu\nu}.
\label{eq:kernel-basis}
\end{equation}
The coefficients contain a tree part and a contribution induced by the one-loop $TJJ$ insertion, $F_i=F_i^{\rm tree}+F_i^{TJJ}$. After the Ward identities are imposed and the on-shell pion projection is performed, Eq.~\eqref{eq:kernel-basis} reduces to the two conserved structures in Eq.~\eqref{eq:gff-decomposition}. In particular,
\begin{align}
A_\pi(q^2)&={f_\pi^2C_Fg_s^2\over N_c}\int\!\dd x\,\dd y\,
\phi_\pi(x)\bigl(F_1+F_3\bigr)\phi_\pi(y),\label{eq:A-projection}\\
D_\pi(q^2)&=-{f_\pi^2C_Fg_s^2\over N_c}\int\!\dd x\,\dd y\,
\phi_\pi(x){2F_5\over q^2}\phi_\pi(y).
\label{eq:D-projection}
\end{align}
The formulas suppress arguments and symmetry-related terms, but make the main projection transparent: the scalar coefficient $F_5$ enters the $D$-term directly.

The collinear denominators contain products such as $x(1-x)y(1-y)$. A leading-twist distribution amplitude vanishes linearly at the endpoints, so the strict leading-power convolutions are finite. At phenomenologically accessible $Q^2$, however, endpoint regions can still carry excessive weight and logarithms from overlapping soft and collinear configurations become important. This motivates the impact-parameter formulation discussed below.

\section{Non-Abelian \texorpdfstring{$TJJ$}{TJJ} correlator and the anomaly}
\label{sec:tjj}
The QCD EMT used in the perturbative vertex is obtained by varying the gauge-fixed action with respect to the metric. The resulting operator contains gauge-invariant quark and gluon pieces together with gauge-fixing and ghost terms. This complete form is required for an off-shell Green function. The renormalized correlator
\begin{equation}
\Gamma^{\mu\nu\alpha\beta}_{ab}(q,p_1,p_2)
=\langle T^{\mu\nu}(q)J_a^\alpha(p_1)J_b^\beta(p_2)\rangle,
\qquad q+p_1+p_2=0,
\label{eq:tjj-def}
\end{equation}
is organized into transverse--traceless, longitudinal or semi-local, and trace sectors. Transverse projectors isolate the physical current components, while the remaining pieces are constrained by diffeomorphism, trace and Slavnov--Taylor identities \cite{Coriano:2024qbr}.

This separation matters because the two gluons entering the pion hard subgraph are off shell. Individual diagrams and individual tensor coefficients may therefore contain gauge-parameter-dependent longitudinal structures. Their dependence is proportional to inverse propagators or to local equation-of-motion terms. When the full vertex is contracted with the leading-twist pion projectors and integrated against the distribution amplitudes, these unphysical pieces cancel. The off-shell decomposition is thus not merely formal: it provides the bookkeeping needed to verify gauge independence of the extracted pion GFFs.

In the massless theory the anomalous trace of the renormalized vertex is controlled by the first beta-function coefficient,
\begin{equation}
\beta(g_s)=-{\beta_0g_s^3\over16\pi^2}+\cdots,
\qquad \beta_0={11\over3}C_A-{2\over3}n_f.
\label{eq:beta0}
\end{equation}
The corresponding scalar form factor has the characteristic behavior
\begin{equation}
\Phi_{\rm anom}(q^2)\sim {\mathcal A\over q^2},
\qquad \mathcal A={g_s^2\over16\pi^2}\beta_0,
\label{eq:anomaly-pole}
\end{equation}
up to convention-dependent signs and tensor normalizations. In an effective action this term is represented by the nonlocal interaction
\begin{equation}
S_{\rm anom}\sim \int\!\dd^4x\,\dd^4y\,
R^{(1)}(x)\,\Box^{-1}(x,y)\,F^a_{\rho\sigma}F^{a\rho\sigma}(y),
\label{eq:nonlocal-action}
\end{equation}
which motivates the dilaton language. The massless propagator in Eq.~\eqref{eq:nonlocal-action} expresses the nonlocality of the anomaly action; it should not be interpreted automatically as an asymptotic elementary particle.

Embedding the vertex in Eq.~\eqref{eq:collinear-factorization} generates an order-$\as^2$ correction to the pion hard amplitude. Its anomaly coefficient is not an adjustable normalization but is fixed by Eq.~\eqref{eq:beta0}. The tensor matching produces an important selection rule. In the combination $F_1^{TJJ}+F_3^{TJJ}$ contributing to $A_\pi$, the isolated beta-function part cancels. The full $TJJ$ vertex can still modify $A_\pi$ through transverse--traceless and other non-anomalous structures. In contrast, the trace coefficient in $F_5^{TJJ}$ survives in $D_\pi$, and the combination $A_\pi+3D_\pi$ in Eq.~\eqref{eq:hard-trace} selects it most directly. The anomaly therefore predicts a hierarchy rather than a common multiplicative correction.

It is also important to distinguish the calculated $TJJ$ insertion from a complete next-order hard kernel. The plotted correction contains the full one-loop $TJJ$ vertex in the selected topology, including its anomalous and non-anomalous pieces, but it does not contain every order-$\as^2$ radiative correction to the pion GFFs. This qualification is particularly relevant for quantitative statements about $D_\pi$, where additional non-anomalous terms may be significant.

\section{Hard-kernel matching and power counting}
\label{sec:matching}
The matching may be understood directly from the virtualities entering the two gluon legs of the $TJJ$ subgraph. In collinear kinematics they scale as
\begin{equation}
p_1^2=-xyQ^2,\qquad
p_2^2=-(1-x)(1-y)Q^2,\qquad q^2=-Q^2.
\label{eq:gluon-virtualities}
\end{equation}
For momentum fractions away from the endpoints, all three invariants are hard and the one-loop vertex is a genuine short-distance object. The scalar functions multiplying its tensor structures contain rational functions of $x$ and $y$, the anomaly coefficient $\mathcal A$, and two- and three-point loop functions. Their renormalized logarithms have the general form $\ln(Q^2/\mu^2)$, $\ln(xyQ^2/\mu^2)$ and $\ln[(1-x)(1-y)Q^2/\mu^2]$. Evolution of the pion distribution amplitude and the running coupling combine with these logarithms so that the residual scale dependence is of higher perturbative order.

The insertion carries one additional loop factor relative to the leading hard exchange. Both the tree amplitude and its $TJJ$ correction have the leading exclusive scaling $f_\pi^2/Q^2$, while the latter is suppressed by an extra power of $\as$ and multiplied in the trace sector by $\beta_0$. Schematically,
\begin{equation}
A_\pi,\ D_\pi\sim {f_\pi^2\over Q^2}
\left[\as(Q)+\as^2(Q)\bigl(c_{\rm reg}+\beta_0c_{\rm anom}\bigr)+\cdots\right].
\label{eq:power-counting}
\end{equation}
The coefficients $c_{\rm reg}$ and $c_{\rm anom}$ denote convolution integrals rather than constants. Equation~\eqref{eq:power-counting} explains why the anomaly is perturbative at large $Q^2$ even though it represents an exact breaking of classical scale invariance. Its normalization is fixed by a Ward identity, but its contribution to a hadronic form factor is filtered by the hard kernel and the pion wave function.

The gauge-cancellation mechanism gives a second useful check. The longitudinal sectors required by the off-shell Slavnov--Taylor identities contain factors proportional to $p_1^\alpha$ or $p_2^\beta$, and terms proportional to the inverse propagators in Eq.~\eqref{eq:gluon-virtualities}. Contraction with the valence pion projectors eliminates the current-longitudinal form factors. A remaining local coefficient can appear temporarily in the scalar projection, but its contribution is antisymmetric or reduces to a total cancellation after convolution over the two pion fractions. Thus the physical $A_\pi$ and $D_\pi$ do not inherit the gauge parameter used to define the internal off-shell vertex.

The separation of quark and gluon EMT operators requires additional care. These operators mix under renormalization, and their individual matrix elements may contain a form factor multiplying $g^{\mu\nu}$ because the separated pieces are not conserved. For the total EMT, conservation removes this apparent third spin-zero form factor. Consequently, a $g^{\mu\nu}$ term found in the intermediate kernel must be matched onto $A_\pi$ and $D_\pi$ rather than interpreted as an independent observable. In the present calculation this matching is precisely what sends the beta-function-controlled coefficient into $D_\pi$ and the trace combination while enforcing its cancellation in $A_\pi$.

Finally, the endpoint domain marks the boundary of the collinear power counting. When $x$ or $y$ becomes parametrically small, one of the virtualities in Eq.~\eqref{eq:gluon-virtualities} is no longer of order $Q^2$, and the nominal hard coefficient samples long-distance physics. A fixed-order calculation then overestimates these configurations even if the distribution amplitude renders the integral mathematically finite. The Sudakov-improved treatment supplies a dynamical restriction to compact partonic configurations and turns this qualitative power-counting requirement into an explicit suppression factor.

\section{Sudakov-improved transverse factorization}
\label{sec:sudakov}
Radiative corrections to an exclusive amplitude generate double logarithms $\as\ln^2(Q^2/k_T^2)$ from overlapping soft and collinear regions. Their resummation is most naturally implemented in impact-parameter space, where $\bm b$ is Fourier conjugate to the partonic transverse momentum. The collinear distribution amplitude is promoted to a transverse-momentum-dependent wave function,
\begin{equation}
\widetilde\Psi_\pi(x,b,\mu)=\int{\dd^2\bm k_T\over(2\pi)^2}
e^{-i\bm k_T\cdot\bm b}\Psi_\pi(x,k_T,\mu),
\label{eq:b-wave-function}
\end{equation}
and the matrix element becomes
\begin{align}
\langle \pi(P_2)|T^{\mu\nu}|\pi(P_1)\rangle
=&{if_\pi^2C_Fg_s^2\over N_c}\int_0^1\!\dd x\,\dd y
\int_0^\infty\!2\pi b\,\dd b\nonumber\\
&\times\widetilde\Psi_\pi(x,b,Q)K^{\mu\nu}(x,y,b,Q)
\widetilde\Psi_\pi(y,b,Q).
\label{eq:b-factorization}
\end{align}
The wave function contains the universal Sudakov exponential
\begin{equation}
\widetilde\Psi_\pi(x,b,Q)=e^{-S(x,b,Q)}\widetilde\Psi_\pi^{(0)}(x,b),
\label{eq:sudakov-wave}
\end{equation}
where, at next-to-leading logarithmic accuracy,
\begin{align}
S(x,b,Q)=&s(x,Q,b)+s(1-x,Q,b)\nonumber\\
&+2\int_{1/b}^{\mu}{\dd\bar\mu\over\bar\mu}\gamma_q(\as(\bar\mu)),\label{eq:sudakov-exponent}\\
s(x,Q,b)=&\int_{1/b}^{xQ}{\dd\bar\mu\over\bar\mu}
\left[\ln{xQ\over\bar\mu}A(\as(\bar\mu))+B(\as(\bar\mu))\right].
\label{eq:s-function}
\end{align}
Here $A$ is governed by the cusp anomalous dimension and $B$ collects single-logarithmic terms. The scale is chosen locally as
\begin{equation}
t=\max\left(\sqrt{xy}\,Q,{1\over b}\right),
\label{eq:hard-scale}
\end{equation}
so that the running coupling is evaluated at the largest perturbative virtuality in the convolution \cite{Sterman:1986aj,Li:1992nu}.

For the nonperturbative input we use a Gaussian transverse profile. At the reference scale $\mu_0$ its impact-parameter representation is
\begin{align}
\widetilde\Psi_\pi^{(0)}(x,b)=&\phi_\pi(x,\mu_0)
\exp\left[-{\beta_\pi^2m_q^2\over x(1-x)}\right]
\exp\left[-{x(1-x)b^2\over4\beta_\pi^2}\right],
\label{eq:gaussian-wave}\\
\phi_\pi(x,\mu_0)=&6x(1-x)\left[1+a_2(\mu_0)C_2^{3/2}(2x-1)\right].
\label{eq:pion-da}
\end{align}
The model combines three complementary suppression mechanisms: the distribution amplitude vanishes at the endpoints, the effective mass term further damps $x\to0,1$, and the Sudakov exponent removes configurations with large $b$ or unresolved soft radiation. For the numerical curves shown below we take $a_2(1\,\mathrm{GeV})=0.2$, $m_q=0.33$ GeV and $\beta_\pi=0.24\,\mathrm{GeV}^{-1}$, following the setup of Ref.~\cite{Coriano:2026gff}.

The Fourier transform of the exchanged-gluon propagator produces modified Bessel functions. In the adopted approximation transverse momentum is retained in that propagator, while intrinsic transverse momenta are neglected in the quark numerators and virtualities. This preserves the dominant non-collinear effect and yields a stable one-dimensional $b$ integral. It is particularly useful for the $TJJ$ kernel, whose logarithmic dependence on the gluon virtualities would otherwise make a fully transverse analytic transform impractical.

The Sudakov factor does not alter the tensor selection rule found in Sect.~\ref{sec:tjj}. It reshapes the convolution and reduces its sensitivity to endpoints, but the isolated anomaly still cancels in $A_\pi$ and survives in the scalar projections. This stability under the transition from collinear to transverse factorization is a useful consistency check.

\section{Numerical results and projection hierarchy}
\label{sec:results}
\begin{figure}[t]
\centering
\includegraphics[width=0.32\textwidth]{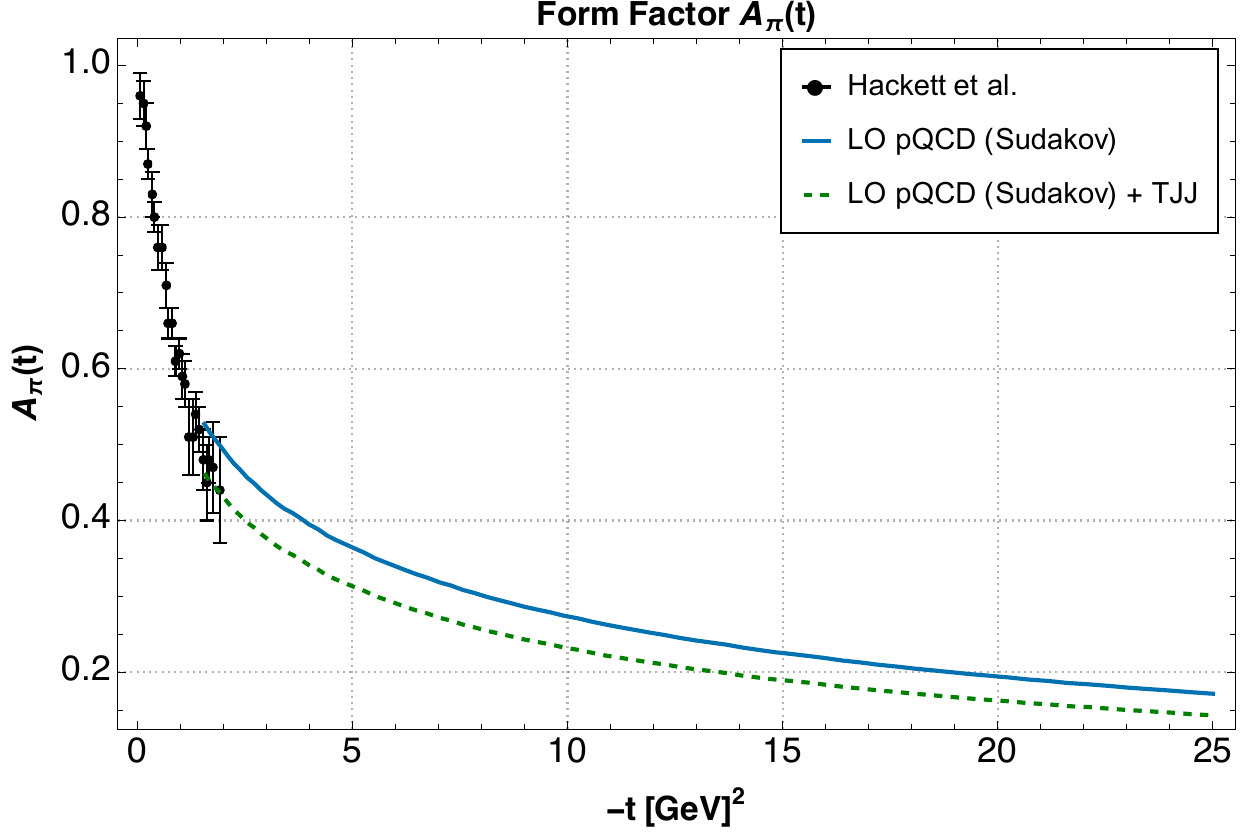}\hfill
\includegraphics[width=0.32\textwidth]{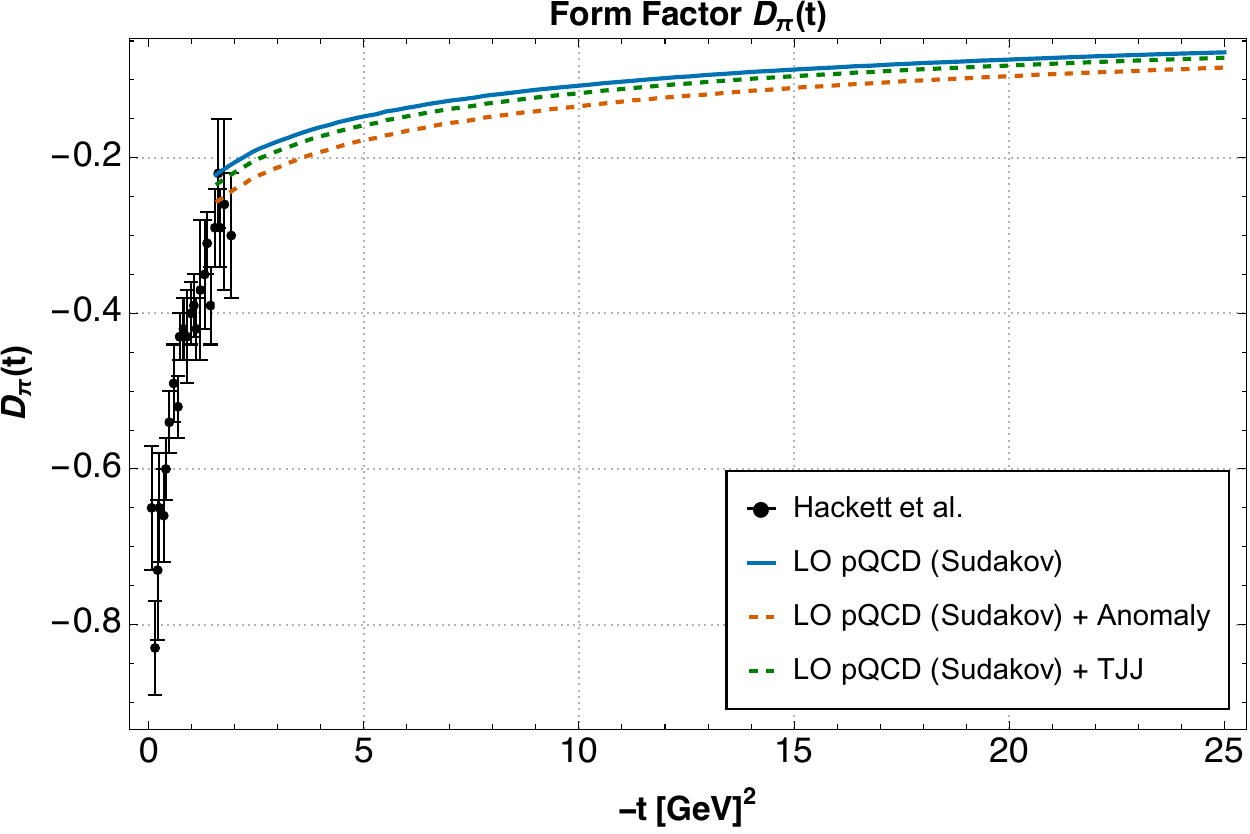}\hfill
\includegraphics[width=0.32\textwidth]{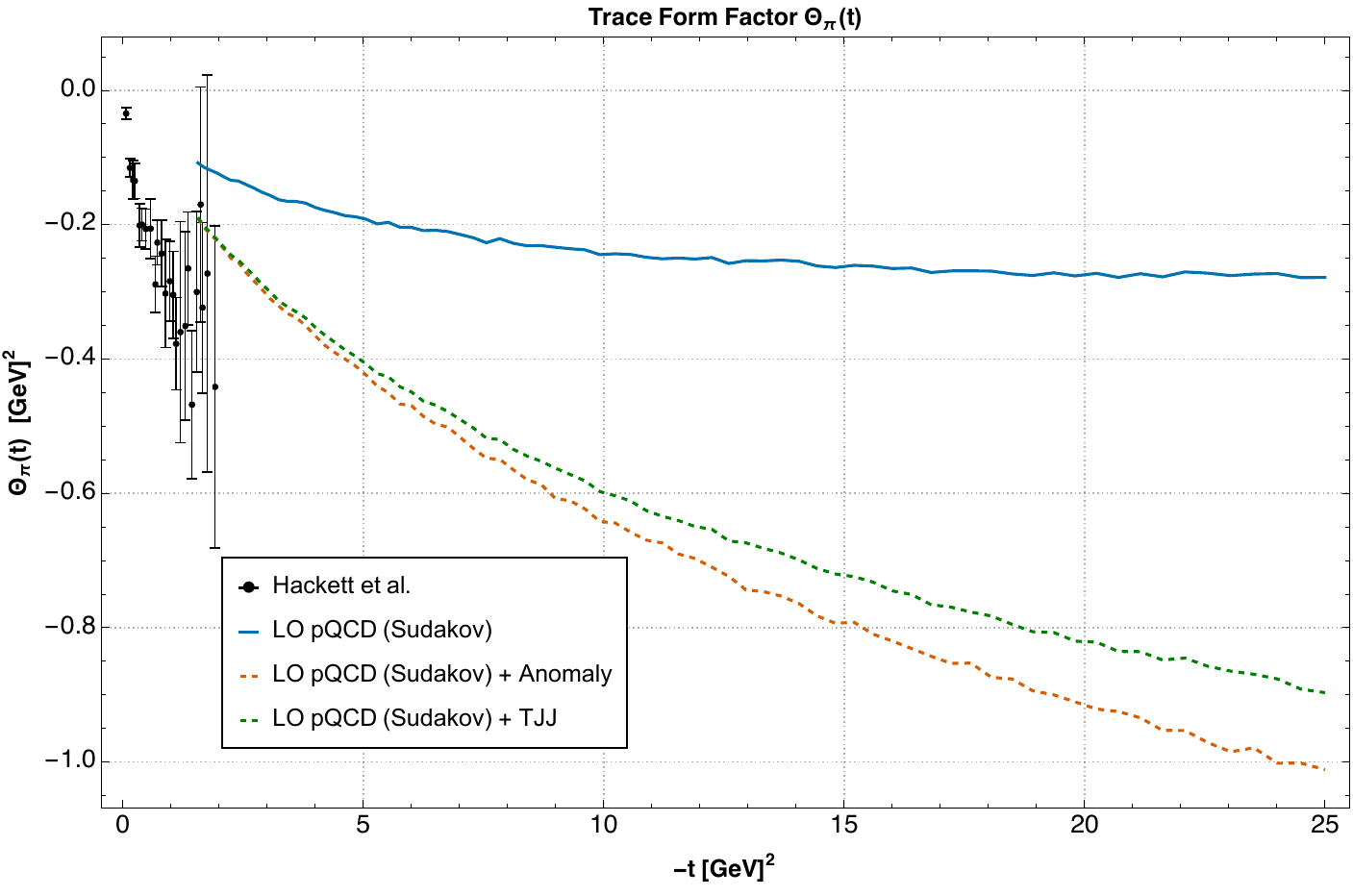}
\caption{Sudakov-improved pion form factors as functions of $Q^2=-t$. From left to right: $A_\pi$, $D_\pi$ and the trace form factor $\Theta_\pi$. The isolated anomaly cancels in $A_\pi$, is important in $D_\pi$, and dominates the $TJJ$ contribution to the trace. Curves and lattice points are reproduced from Ref.~\cite{Coriano:2026gff}; model parameters are given below Eq.~\eqref{eq:pion-da}.}
\label{fig:results}
\end{figure}

Figure~\ref{fig:results} compares the leading Sudakov prediction, the isolated anomaly contribution and the result including the full $TJJ$ insertion with lattice-QCD determinations \cite{Hackett:2023pyn}. The comparison reaches into a low and intermediate momentum region in which strict hard factorization is not expected to be quantitatively complete. It should therefore be read as a study of the direction, relative size and projection pattern of the anomaly correction, not as a precision low-energy fit.

The momentum form factor $A_\pi$ exhibits the cleanest cancellation. There is no separate anomaly curve because the beta-function-controlled terms cancel in its leading-twist projection. The complete $TJJ$ insertion nevertheless lowers the leading Sudakov result at smaller $Q^2$. This shift comes from the non-anomalous tensor structures of the same renormalized vertex and locally improves the overlap with the lattice points.

The behavior of $D_\pi$ is different. The scalar part of the $TJJ$ vertex projects efficiently onto $F_5$ and gives a sizable anomaly contribution. This is physically natural because the $D$-term controls the stress distribution and is sensitive to scalar dynamics. The full $TJJ$ curve should nevertheless not be mistaken for a complete order-$\as^2$ prediction: other radiative pieces may change both its normalization and shape.

The trace form factor provides the sharpest signature. Through Eq.~\eqref{eq:hard-trace}, it directly selects the scalar combination of $A_\pi$ and $D_\pi$. The anomaly dominates the displayed $TJJ$ correction across the momentum range considered. A universal normalization shift could not produce this pattern; the different response of the three projections is evidence that the trace component has been isolated consistently.

At low $Q^2$, soft overlap, higher-twist pion amplitudes, explicit chiral symmetry breaking and nonperturbative scalar dynamics become increasingly important. Instanton-vacuum calculations, for example, supply a complementary soft and semi-hard description of pressure and shear distributions \cite{Liu:2024zahed}. A broader phenomenology should match such contributions to the perturbative $TJJ$-improved kernel while avoiding double counting in the scalar trace channel.

\section{Dilaton channel and anomaly sum rule}
\label{sec:dilaton}
The anomaly pole also admits a dispersive interpretation. For general virtualities and a nonzero fermion mass, its scalar form factor can be written as
\begin{equation}
\Phi_{\rm anom}(q^2;p_1^2,p_2^2,m^2)
={1\over\pi}\int_0^\infty\!\dd s\,
{\rho_{\rm anom}(s;p_1^2,p_2^2,m^2)\over s-q^2-i0},
\label{eq:dispersion}
\end{equation}
with a finite sum rule
\begin{equation}
{1\over\pi}\int_0^\infty\!\dd s\,\rho_{\rm anom}(s;p_1^2,p_2^2,m^2)=\mathcal A.
\label{eq:sum-rule}
\end{equation}
The spectral density changes when masses and virtualities are varied, but its total area is fixed by the anomaly coefficient. In the conformal limit the spectral weight flows toward the origin,
\begin{equation}
\rho_{\rm anom}(s;m^2)\longrightarrow \pi\mathcal A\,\delta(s),
\qquad m\longrightarrow0,
\label{eq:spectral-flow}
\end{equation}
and Eq.~\eqref{eq:dispersion} reduces to the pole in Eq.~\eqref{eq:anomaly-pole} \cite{Coriano:2025sumrule}.

The intermediate states contributing to $\rho_{\rm anom}$ are quark pairs and, in the non-Abelian sector, gluonic and ghost states. The trace projection organizes them into a color-singlet scalar channel. This is the sense in which the perturbative dilaton can be regarded as a correlated pair exchange. The anomaly sum rule constrains the scalar part of the hard kernel before convolution; it is not a separate normalization condition for each hadronic GFF. The pion wave functions and the tensor projectors redistribute the fixed scalar strength among observable form factors, allowing it to cancel in $A_\pi$ while remaining visible in $D_\pi$ and $\Theta_\pi$.

There is a useful conceptual distinction between an anomaly pole and a genuine particle pole. Away from the massless on-shell light-cone configuration, the residue of $q^2\langle TJJ\rangle$ at $q^2=0$ vanishes. A nonzero particle-like residue is obtained only for massless internal fields and transverse on-shell external gauge bosons \cite{Armillis:2009pq,Armillis:2010qk}. Moreover, the light-cone vertex contains both a scalar anomalous pole and an independent traceless pole structure. Only the former is fixed by the trace Ward identity and belongs to the dilaton channel. This distinction is essential when a $TJJ$ vertex is inserted into an off-shell hard subgraph.

At hadronic scales the same conserved anomaly strength is expected to be redistributed over physical scalar intermediate states, including pion continua, scalar resonances and gluonic components. The perturbative pole is therefore best viewed as the short-distance representative of a broader scalar spectral channel. The matching between this short-distance description and a hadronic saturation of the trace channel remains an important open problem.

\section{Conclusions}
\label{sec:conclusions}
The pion GFFs provide a clean arena in which hard exclusive factorization, EMT Ward identities and the QCD conformal anomaly meet. Inserting the renormalized non-Abelian $TJJ$ correlator into the pion hard kernel produces a calculable scalar correction whose normalization is fixed by $\beta_0$. The off-shell tensor decomposition, including its gauge-fixing and semi-local sectors, is necessary to establish a gauge-independent on-shell pion amplitude.

The resulting projection hierarchy is the central result. The isolated anomaly cancels in the leading-twist momentum form factor $A_\pi$, contributes importantly to the mechanical form factor $D_\pi$, and is selected most directly by the trace form factor. Transverse-momentum dependence and Sudakov resummation regulate endpoint and large-$b$ regions without changing this hierarchy. Numerically, the full $TJJ$ insertion moves the perturbative curves toward lattice results in part of the accessible range, although a complete low-energy or full order-$\as^2$ description requires additional contributions.

The dispersive sum rule further clarifies the dilaton interpretation. The anomaly pole is the conformal-limit localization of a correlated scalar spectral density, not an independently postulated particle. Its fixed strength is matched through the pion wave function and distributed non-universally among $A_\pi$, $D_\pi$ and $\Theta_\pi$. Extending this matching to a consistent combination of perturbative, lattice and nonperturbative hadronic information is a natural next step toward a quantitative picture of pion mechanical structure.

\section*{Acknowledgements}
We thank Hsiang-nan Li for collaborating on this project. This work is partially supported by INFN under Iniziativa Specifica QG-sky and by the National Centre for HPC, Big Data and Quantum Computing.

\end{document}